\documentclass[aps,prd,reprint,superscriptaddress,nofootinbib,floatfix,longbibliography]{revtex4-2}

\usepackage[dvipsnames]{xcolor}
\usepackage{amssymb,amsmath,amsfonts,bm,slashed,soul,ragged2e,graphicx,epstopdf,hyperref,array,float}
\usepackage[utf8]{inputenc}
\usepackage{colortbl,color,caption,subcaption}
\enlargethispage{\baselineskip}
\definecolor{steelblue}{RGB}{25,25,112}
\definecolor{dullblue}{rgb}{0,0.298,0.49}
\definecolor{darkred}{rgb}{0.545,0,0}
\definecolor{darkorange}{RGB}{222,132,69}
\definecolor{darkgreen}{RGB}{126,171,85}
\definecolor{blue2}{cmyk}{1, 0.1, 0.1, 0}
\hypersetup{colorlinks,linkcolor={darkred},citecolor={blue},urlcolor={dullblue}}
\DeclareCaptionJustification{justified}{\justifying}
\everymath{\displaystyle} 

\usepackage{titlesec}
\usepackage{mathtools}

\newcommand{\be}{\begin{equation}}
\newcommand{\ee}{\end{equation}}
\def \d {{\rm d}}
\usepackage{physics}
\DeclareMathOperator\diag{diag}

\usepackage{chemfig,siunitx}

\begin{document}

\title{Dynamical Lensing Tomography of Black Hole Ringdown}

\author{Zhen Zhong}
\email{zhen.zhong@tecnico.ulisboa.pt}
\affiliation{CENTRA, Departamento de F\'{\i}sica, Instituto Superior T\'ecnico -- IST, Universidade de Lisboa -- UL,
Avenida Rovisco Pais 1, 1049-001 Lisboa, Portugal}
\author{Vitor Cardoso}
\email{vitor.cardoso@nbi.ku.dk}
\affiliation{Center of Gravity, Niels Bohr Institute, Blegdamsvej 17, 2100 Copenhagen, Denmark}
\affiliation{CENTRA, Departamento de F\'{\i}sica, Instituto Superior T\'ecnico -- IST, Universidade de Lisboa -- UL,
Avenida Rovisco Pais 1, 1049-001 Lisboa, Portugal}
\author{Yifan Chen}
\email{yifan.chen@nbi.ku.dk}
\affiliation{Center of Gravity, Niels Bohr Institute, Blegdamsvej 17, 2100 Copenhagen, Denmark}
\begin{abstract}
Strong gravitational lensing occurs when photons pass through the vicinity of a black hole. We investigate this phenomenon in the context of a gravitational-wave event, specifically when a black hole is settling into its final state. The deflection angle of photons mimics the ringdown pattern of the gravitational wave at intermediate times. At late times it has an inverse cubic dependence on observation time. The deviation angle increases exponentially as photons approach the photon ring orbit, reflecting its unstable nature. Our findings are directly applicable to imaging scenarios involving stars against the background of compact binaries, or circumbinary accretion disks, particularly during the merger of two black holes.
\end{abstract}
\maketitle

\section{Introduction}
One of the foundational tests of General Relativity concerns the deflection of light by massive bodies~\cite{Will:2014kxa,Lemos:2019bmc,Bartelmann:1999yn,Bartelmann:2010fz}: the path of light rays is ``bent'' by mass-energy, which distorts the spacetime fabric. In the last decade, it has become clear that the universe is flooded with traveling spacetime distortions, gravitational waves (GWs) traveling at the speed of light. Gravitational-wave astronomy provides a wealth of information on compact objects, particularly on the behavior of gravity in highly dynamical, high-curvature and redshift setups.

An intriguing prospect concerns the detectability of {\it light} bending by GWs, which would add an extra information channel on strong gravity. This possibility was raised and studied by different authors in the weak field regime~\cite{1993A&A...268..823L,1994ApJ...426...74F,PhysRevLett.72.3301,Kaiser:1996wk,Damour:1998jm}. Our purpose here is to study light deflection near compact objects in highly dynamic spacetimes. In particular, we will study the appearance of a black hole (BH) which is approaching a quiescent state via relaxation in its characteristic modes of oscillation, called quasinormal modes (QNMs). The relaxation stage is also referred to as ringdown stage, and is a generic late-time description of the coalescence of compact objects~\cite{Berti:2009kk,Baibhav:2023clw}. Our results could describe stars on the background of a binary BH merger, or the appearance of a circumbinary accretion disk surrounding two merging compact objects. 

We work in units where $G = c = 1$, and adopt the metric convention $(-,+,+,+)$.

\section{Ray Tracing towards a Relaxing Black Hole}
A relaxing BH is well described by general relativistic first-order perturbation theory, where spacetime is expressed by small fluctuations away from the final, stationary state. Uniqueness results in General Relativity suggest that the final state, to a good approximation, is a Kerr BH~\cite{Chrusciel:2012jk,Cardoso:2016ryw,Cardoso:2021wlq}. We thus take our relaxing BH spacetime, of mass $M$ and angular momentum $a M$ to be given by 
\begin{equation}
g_{\mu\nu} = g_{\mu\nu}^{\mathrm{(0)}} + \,h_{\mu\nu} \, ,
\end{equation}
where $g_{\mu\nu}^{\mathrm{(0)}}$ is the Kerr metric (we will use Boyer-Lindquist $(t,r,\theta,\phi)$ coordinates throughout~\cite{Bardeen:1972fi}). The term $h_{\mu\nu}$ denotes first-order perturbations.

Instead of working directly with the metric perturbations $h_{\mu\nu}$, we consider instead curvature perturbations~\cite{Teukolsky:1973ha,Teukolsky:1972my}. These can be separated and decoupled and all the relevant information about radiative degrees of freedom is encapsulated in a master variable
\begin{equation}
\psi_{s}(t, r, \theta, \phi) = \frac{{}_{s} Z }{\sqrt{2\pi}} 
\, {}_{s} S(\theta) \, {}_{s}R(r) e^{-i \omega t + i m \phi} \,. 
\end{equation}
The gravitational quantity of interest is $\Psi_4^{(1)}=\psi_{-2}(t, r, \theta, \phi)(r-ia\cos\theta)^{-4}$, the first-order perturbation of one of the Weyl scalars, which is directly related to the amplitude of GWs far from the source.
Notation and conventions are shown in Supplemental Material~A.1~\cite{SupplementalMaterial}. The function ${}_{s} S (\theta)$ is a spin-weighted spheroidal harmonic, while the radial Teukolsky function ${}_{s}R(r)$ obeys a homogeneous second-order differential equation. We fix its normalization by calculating energy fluxes at infinity, and comparing to numerical relativity results. Then, ${}_{s}Z$ controls the amplitude of the GW, by matching the ringdown stage to some physically generated process. The characteristic modes are found by imposing proper boundary conditions on ${}_{s}R(r)$ , which selects only a discrete set of frequencies $\omega$, the ringdown frequencies~\cite{Berti:2009kk}. All the functions and quantities carry a dependence on angular numbers $\ell, m$ and a possible overtone number $n$~\cite{Teukolsky:1973ha,Teukolsky:1972my,Berti:2009kk}. For simplicity, we do not include these indices in the subscripts. A GW signal comprises a superposition of various modes, but the $\ell=|m|=2$ dominates (so far it is the only one detected beyond any doubt). Additionally, we exclusively address the $n=0$ mode, which lasts longer compared to the more rapidly decaying higher overtones. For similar reasons we neglect nonlinearities, which are of much smaller amplitude even for equal-mass mergers, and of shorter duration~\cite{Cheung:2022rbm,Redondo-Yuste:2023seq}. 
For QNMs characterized by frequency $\omega$ and labeled by $(\ell, m, n)$, there exists a corresponding ``mirror'' mode with frequency $-\bar{\omega}$ and $(\ell, -m, n)$. In the final spin frame, these paired modes primarily represent GW emissions propagating in opposite directions along the spin axis~\cite{Li:2021wgz, Berti:2005ys, Cook:2014cta, Krivan:1997hc}. The observed waveform typically manifests as a superposition of these two distinct sets of modes. However, one of the paired modes may be less prominent due to weaker excitation~\cite{Berti:2005ys, Hamilton:2023znn}. In the following analysis, we primarily focus on the $m=2$ mode with $\mathrm{Re}(\omega) > 0$, which predominantly emits toward the southern hemisphere, as more general cases exhibit similar features, as discussed in Supplemental Material~C~\cite{SupplementalMaterial}.

Once ${}_{-2}R(r)$ is known, we reconstruct the metric fluctuation $h_{\mu\nu}$ following well-known prescriptions~\cite{Chrzanowski:1975wv, Cohen:1974cm, Kegeles:1979an}, summarized in Supplemental Material~A.2~\cite{SupplementalMaterial}. We initiate the ringdown at $t = r_*$, where $r_*$ denotes the tortoise radial coordinate, by applying a Heaviside-like filter to the perturbation, $h_{\mu\nu} \to H(t - r_*)\, h_{\mu\nu}$, along the light cone, with
\begin{equation}
H(t - r_*) = \frac{1}{2} \left(\tanh{\frac{t - r_*}{\chi}} + 1\right) \,.
\end{equation}
In the following, we employ a relatively sharp truncation in the filter function $H(t - r_*)$, with $-\chi \mathrm{Im}(\omega) = 0.1$.

During the ringdown phase, a photon trajectory is influenced by the metric perturbations encountered along its path, governed by the geodesic equation:
\begin{equation}
\dv[2]{x^\mu}{\tau}+\Gamma^\mu{}_{\nu \rho} \dv{x^\nu}{\tau} \dv{x^\rho}{\tau}=0 \,, \label{eq:geodesic}
\end{equation}
where $\Gamma$ denotes the Levi-Civita connection, and $\tau$ represents the proper time.

\begin{figure}[htbp]
 \includegraphics[width=\linewidth]{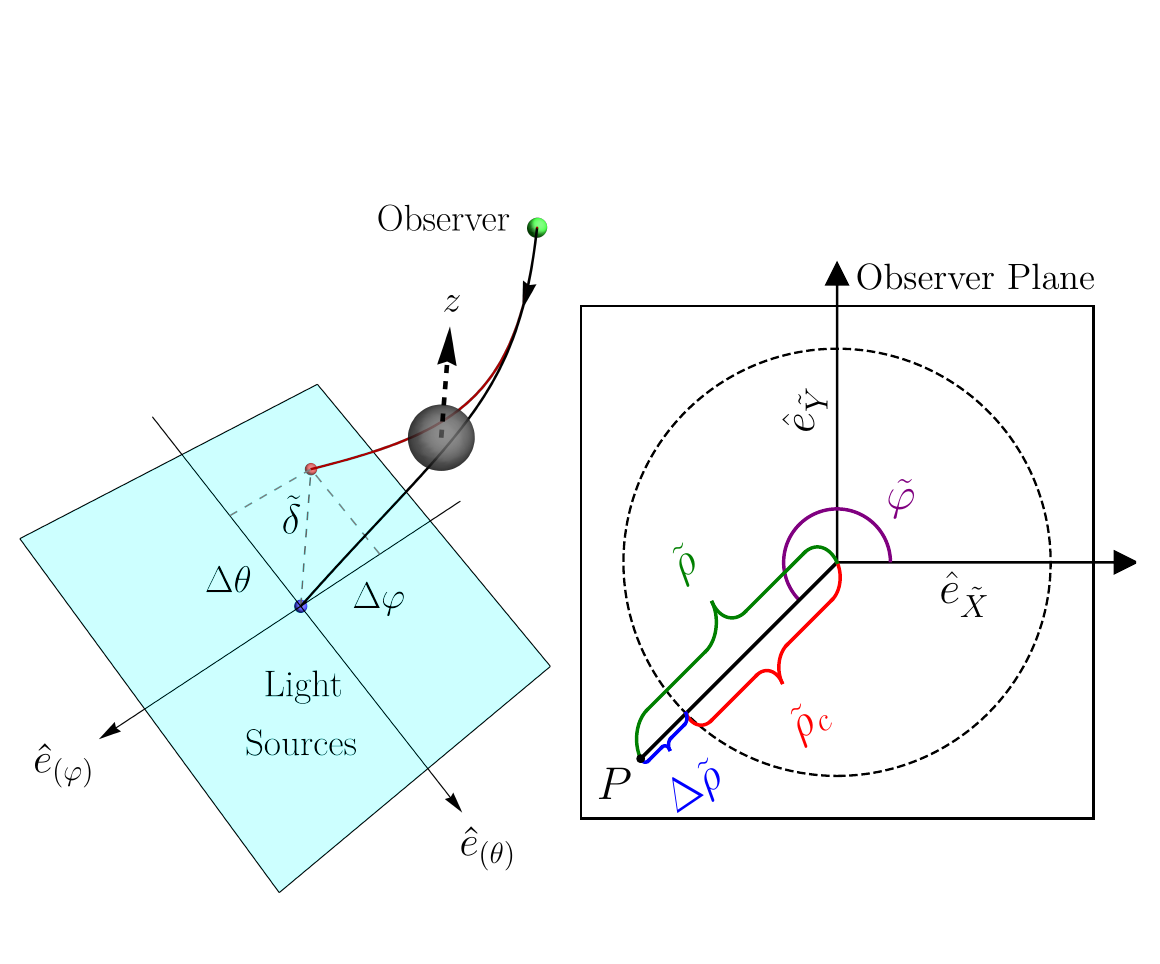}
\caption{Schematic diagram of the coordinate system. \textbf{Left:} Illustration of a light ray originating from the observer (green point) and traced backward (black arrow) toward the BH, terminating at the celestial sphere plane at infinity (blue plane). The black and red lines represent the geodesics in unperturbed Kerr spacetime and those deflected by metric perturbations, respectively. The grand arc difference is denoted by $\tilde{\delta}$, with angular variations in the azimuthal and polar directions labeled as $\Delta \phi$ and $\Delta \theta$. 
\textbf{Right:} Diagram of the observer plane with polar coordinates $(\tilde{\rho}, \tilde{\varphi})$. The critical curve $\tilde{\rho}_c(\tilde{\varphi})$ is shown as a black contour, and the relative impact parameter is defined as $\Delta \tilde{\rho} \equiv \tilde{\rho} - \tilde{\rho}_c$.}
\end{figure}

To characterize the observable effects, it is essential to define the observer frame. In an unperturbed Kerr BH scenario, a zero angular momentum observer (ZAMO) frame is typically used~\cite{Frolov:1998wf}. However, in a dynamic spacetime, the ZAMO tetrads are neither orthogonal nor normalized. To rectify this, we apply a Gram-Schmidt orthogonalization process to establish a normalized, orthogonal tetrad basis derived from the ZAMO tetrads, as detailed in Supplemental Material~B.1~\cite{SupplementalMaterial}. Using this newly defined observer frame, we initiate geodesics towards the BH employing backward ray tracing. Here, each pixel on the observer plane, represented by coordinates $(\tilde{X}, \tilde{Y})$~\cite{Bardeen:1973tla,Gralla:2017ufe,Gralla:2019drh}, corresponds to photon momentum in this frame. The photon's $4$-momentum in the dynamic spacetime $p^\mu \equiv \d x^\mu/\d \tau$ is then constructed from this orthogonal tetrad basis, similar to the procedure established in Kerr spacetime~\cite{Cunha:2016bpi}.
For convenience, we also introduce polar coordinates for the observer plane centered on the direction directly facing the BH in Kerr spacetime, denoted as $(\tilde{\rho}, \tilde{\varphi})$, where $\tilde{\rho}$ is the impact parameter and $\tilde{\varphi}$ is the angle.

Photon geodesics in the relaxing Kerr spacetime are then computed. We position an observer at $r_o = 10^9\,M$, confirming convergence of results at large $r_o$ due to the decay of $h_{\mu\nu}$ amplitude as $1/r$ at greater distances. Notice that even though we employ backward ray tracing, photon propagation in the $+t$ direction still aims towards the observer, aligning with the GWs that co-propagate in the same direction upon departure from the BH.

\section{Lensing Tomography}
\begin{figure}[htbp]
    \centering
    \includegraphics[width=\linewidth]{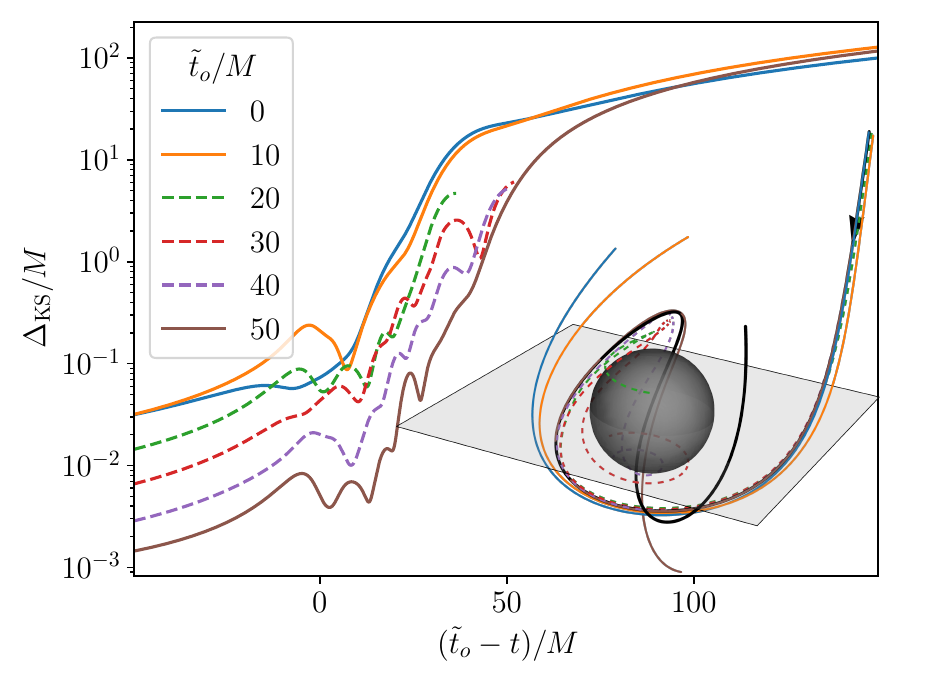}
    \caption{Growth of geodesic deviation $\Delta_{\rm KS}$, representing the distance between unperturbed and deflected geodesics at the same backward-propagating time $\tilde{t}_o - t$, in a relaxing Kerr BH spacetime with spin $a = 0.7M$, viewed from a face-on observer with $\theta_0 \approx \pi$, is shown in Cartesian KS coordinates. Different colors represent geodesics launched at various observation times $\tilde{t}_o$, as detailed in the accompanying inset panel. The dashed line marks a trajectory terminating on the BH, whereas the others extend towards infinity. The metric perturbation is modeled using the $n = 0$ and $\ell = m = 2$ mode, with the magnitude normalized to $|{}_{-2}Z| = 0.24$. The initial impact parameters are set at $\tilde{\rho} = 5.041M$, slightly exceeding the critical curve $\tilde{\rho}_c = 5.04M$ in Kerr spacetime. {\bf Inset:} Illustration of the corresponding photon trajectories. The black line represents unperturbed geodesics in Kerr spacetime, with an arrow indicating the direction from the observer to the BH to depict the backward ray-tracing method used. The gray plane indicates the BH equator. }
    \label{fig:traj}
\end{figure}

We explore geodesics in a relaxing Kerr spacetime, particularly focusing on the merger events of supermassive binary BHs (SMBHBs) with equal mass ratios. The remnant BH from each mass ratio merger events typically characterized by an angular momentum parameter $a \sim 0.7\,M$~\cite{Buonanno:2006ui}, a value consistent with numerous observations~\cite{Nitz:2018imz, LIGOScientific:2018mvr} and considered as our fiducial value. The GW signal from these mergers is dominated by the $\ell=|m|=2$ mode~\cite{Berti:2007fi, Berti:2009kk,Berti:2005ys,maggiore2018gravitational}, with the $n=0$ mode persisting the longest. This fundamental mode is characterized by $M\omega=0.5326-0.08079i$~\cite{Berti:2009kk,Stein:2019mop}.

In the vicinity of the BH, the metric perturbation amplitude can reach up to $M^2\Psi_{4}^{(1)} \sim 0.1$~\cite{Berti:2007fi}. The total radiated energy during the ringdown, $E_{\mathrm{tot}}$, is limited to approximately $3\%$ of the remnant BH's mass~\cite{Berti:2007fi}, setting an upper limit on the magnitude of $|{}_{-2}Z|$ for the $\ell=m=2$ mode at about $0.24$. We adopt this upper bound as the fiducial value for our analyses, our results can be trivially re-scaled for generic amplitudes.

We first consider a face-on observer with $\theta_o \approx \pi$. Figure~\ref{fig:traj} illustrates backward photon trajectories, all launched from the same direction, using identical backward impact parameters relative to the observer positioned at $r_0=10^9M$. These trajectories are depicted using the Cartesian Kerr–Schild (KS) coordinate system $(x, y, z)$.
The black line represents a geodesic in an unperturbed Kerr spacetime, while the various colors indicate different observation times $\tilde{t}_o \equiv t_o - r_*(r_o)$, where $\tilde{t}_o = 0$ is approximately the time when the GW ringdown initially reaches the observer. The bottom panel displays the growth of deviations between unperturbed and deflected geodesics, defined as $\Delta_{\rm KS} \equiv (\Delta x^2 + \Delta y^2 + \Delta z^2)^{1/2}$, representing the distance between the two trajectories at the same backward propagating time $\tilde{t}_o - t$, colors corresponding to trajectory lines in the inset panel.

In Fig.~\ref{fig:traj}, we select $\tilde{\rho} = 5.041M$ as the (backward) impact parameter, which is close to the critical curve $\tilde{\rho}_c \approx 5.04 M$ for this BH spin. This curve, defined in Kerr spacetime, represents the geodesics that can propagate around the BH indefinitely. Some of the perturbed geodesics, shown in dashed lines, terminate on the BH instead of propagating towards infinity, indicating a distortion of the critical curve during the ringdown.

The evolution of geodesic deviations, launched at different observer times, initially shows a weak increase after departing from the observer. Geodesics launched later exhibit deviations exponentially smaller than those launched earlier, reflecting the exponential decay of GW amplitude near the BH during the ringdown phase. Closer to the BH, all deviations begin to exhibit exponential growth due to the instability of the radial potential in the light ring region~\cite{1965SvA.....8..868P,1968ApJ...151..659A,Luminet:1979nyg,Campbell:1973ys,1972ApJ...173L.137C,Falcke:1999pj,MTB,Cardoso:2008bp,Johnson:2019ljv,Gralla:2019drh,Cardoso:2021sip,Chen:2022kzv}. Here, the interplay between light ring instability and GW induces additional oscillation features in the exponential growth. The geodesics that depart from the light ring region begin to develop deviations that increase linearly with time.

To quantify the lensing during ringdown, we analyze backward photons that ultimately reach infinity, corresponding to far-away sources. We assess the angular difference between the asymptotic direction for Kerr geodesic and the deflected geodesic as observables, including the grand arc difference $\tilde{\delta}$, and variations in the azimuthal angle $\Delta \phi$ and polar angle $\Delta \theta$.

\begin{figure}[htbp]
    \centering
    \includegraphics[width=\linewidth]{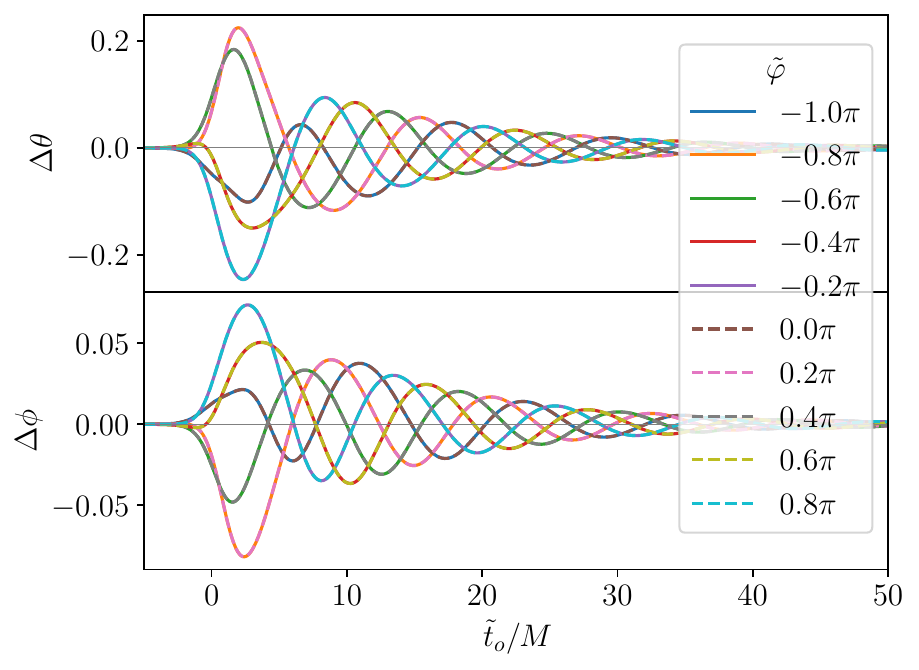}
   \caption{Evolutions of the deflected angles in the azimuthal ($\Delta \phi$) and polar ($\Delta \theta$) components for geodesics terminating at infinity, plotted as a function of observation time $\tilde{t}_o$. The configuration of the BH, observer, and ringdown is consistent with the setup described in Fig.~\ref{fig:traj}. Different colors illustrate geodesics launched from various angles on the polar coordinate of the observer’s plane $\tilde{\varphi}$, all with fixed impact parameters of $\tilde{\rho}=6M$. The symmetrical patterns observed in cases separated by $\pi$ in $\tilde{\varphi}$ highlight the $m=2$ symmetry of the BH ringdown.}
    \label{fig:phi}
\end{figure}
Figure~\ref{fig:phi} displays the deflected angles in polar $(\Delta \theta)$ and azimuthal $(\Delta \phi)$ components as a function of observer frame time, $\tilde{t}_o$, for trajectories initially launched from various polar coordinate angles $\tilde{\varphi}$ on the observer plane. The impact parameters are fixed at $\tilde{\rho}= 6M$. Due to ringdown metric perturbations, axial symmetry is not preserved even in face-on observations. Notably, both $\Delta \phi$ and $\Delta \theta$ show identical variations for values of $\tilde{\varphi}$ separated by $\pi$. This pattern is precisely attributable to the $m=2$ mode of the metric perturbations, akin to those generated by vector or tensor superradiant clouds~\cite{Chen:2022kzv}, demonstrating that the deflected angles can effectively dissect the polarization pattern of the GWs around the BH.

\begin{figure}[htbp]
    \centering
    \includegraphics[width=\linewidth]{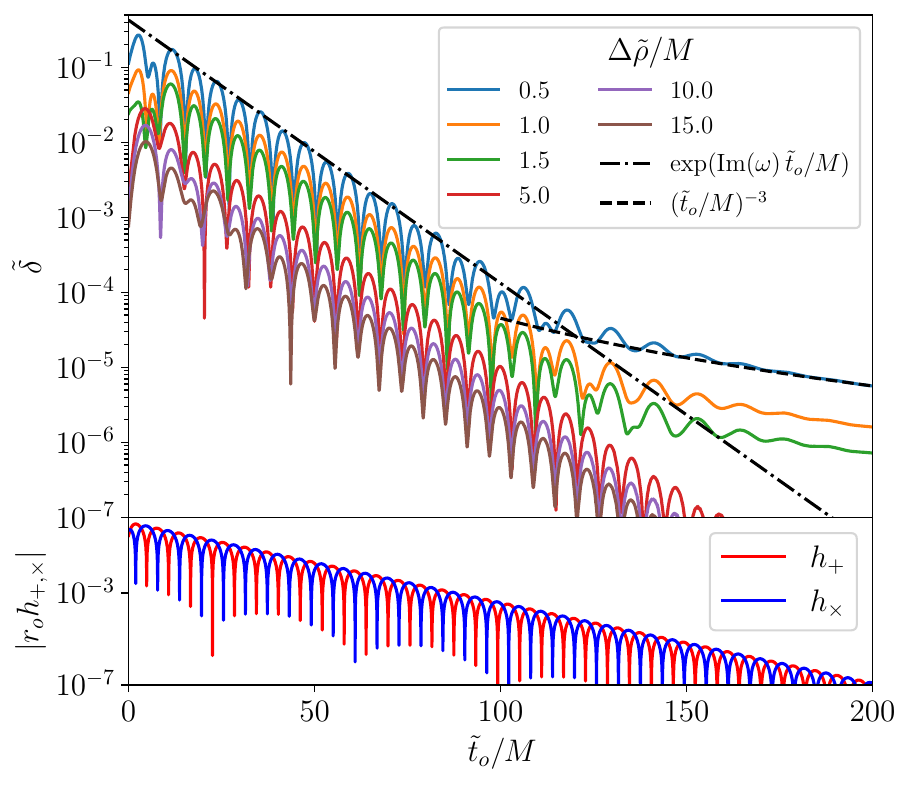}
    \caption{{\bf Top:} Deflection angle of geodesics terminating at infinity, $\tilde{\delta}$, defined as the grand arc difference from the unperturbed direction, plotted as a function of observation time $\tilde{t}_o$. Different colors represent geodesics launched with varying relative impact parameters $\Delta \tilde{\rho}$ along the $+\tilde{X}$-axis. The BH, observer, and ringdown configuration are consistent with the setup described in Fig.~\ref{fig:traj}. Early stages show a dominant exponential decay and oscillation behavior, in very good agreement with the GW ringdown depicted in the bottom panel. The envelope of the $\Delta\tilde{\rho}/M = 0.5$ curve is fitted by $\tilde{\delta} \sim A_{\tilde{\delta}} \exp(\mathrm{Im}(\omega) \, \tilde{t}_o)$, illustrated by the dot-dashed line. At later times, a slow decay appears, fitted by $\tilde{\delta} \propto \tilde{t}_o^{-3}$, shown by the dashed line. {\bf Bottom:} Amplitudes of the GWs in the two polarizations, $h_+$ and $h_\times$, in transverse traceless gauge, as observed by an observer at $\theta_o / \pi = 1 - 10^{-4}$.}
    \label{fig:rho}
\end{figure}

In the top panel of Fig.~\ref{fig:rho}, we depict the evolution of the deflection angle $\tilde{\delta}$ for different relative impact parameters $\Delta \tilde{\rho} \equiv \tilde{\rho} - \tilde{\rho}_c$, where $\tilde{\rho}_c \approx 5.04M$ is defined as the critical curve in Kerr spacetime (for $a=0.7M$). The early time evolutions of $\tilde{\delta}$ consistently exhibit a pattern of exponential decay and oscillation that aligns with the time-dependence of the gravitational waveform shown in the bottom panel. This alignment suggests that these observables effectively capture key characteristics of the BH ringdown. We fit the envelope of the $\Delta\tilde{\rho}/M=0.5$ case using the ringdown decay rate, specifically, $\tilde{\delta} \sim A_{\tilde{\delta}} \exp(\mathrm{Im}(\omega) \, \tilde{t}_o)$, as shown in the dot-dashed line. This demonstrates a clear overlap in the early stages. The amplitude of the deflection angle $A_{\tilde{\delta}}$ is well described at large $\Delta \tilde{\rho}$ by
\begin{equation}
A_{\tilde{\delta}}  \sim \frac{1.2 \sqrt{E_{\mathrm{tot}}/M}}{\Delta \tilde{\rho} / M} \,, \label{eq:delta_E_rho}
\end{equation}
with $E_{\mathrm{tot}}/M = 0.03$ in our fiducial case.

Note that the scaling $A_{\tilde{\delta}} \propto 1/\Delta \tilde{\rho}$ differs from the inverse-cubic dependence reported in Refs.~\cite{Damour:1998jm,Kopeikin:1998ts,Kopeikin:1999ev,Kopeikin:2006gf}. This distinction arises because, in our case, the dominant deflection occurs locally when the photon first encounters the ringdown perturbation near $t \approx r_*$ -- close to the BH and at early times -- when the photon and GW propagation directions are not yet fully aligned. As a result, the deflection, which is proportional to the contraction $h_{\mu\nu} p^\mu p^\nu$, is primarily governed by the $1/r$ component of the metric perturbation, rather than the $1/r^3$ contribution. In contrast, the inverse-cubic scaling in Refs.~\cite{Damour:1998jm,Kopeikin:1998ts,Kopeikin:1999ev,Kopeikin:2006gf} assumes a source that emits GWs continuously and nearly isotropically long before the photon arrives, such that only the component of the GW field with momentum aligned with the photon direction contributes appreciably to the net deflection.

Furthermore, as the impact parameter approaches the critical curve, specifically when $\Delta \tilde{\rho} \ll M$, there is a noticeable increase in the overall amplitude of $\tilde{\delta}$. This enhancement can be attributed to the instability of the BH light ring region, which causes an exponential growth in the perturbed deviation during propagation around the BH~\cite{1965SvA.....8..868P,1968ApJ...151..659A,Luminet:1979nyg,Campbell:1973ys,1972ApJ...173L.137C,Falcke:1999pj,MTB,Cardoso:2008bp,Johnson:2019ljv,Gralla:2019drh,Chen:2022kzv}.

On the other hand, at late times, the evolution exhibits a power-law decay. This distinction arises from deflections developing in two distinct regions. During the early stages, the photon encounters the GW field close to the BH, where the $1/r$ component of the perturbation dominates and induces a deflection that decays locally in an exponential fashion—reflecting the ringdown behavior of the GW itself. At late times, the encounter occurs farther from the BH, where the photon and GW propagate nearly anti-parallel. In this regime, the deflection effect receives non-vanishing contributions only from the $1/r^3$ component of the metric perturbations. Indeed, our results are consistent with a power-law scaling, $\tilde{\delta} \propto \tilde{t}_o^{-3}$, as shown by the dashed fit in Fig.~\ref{fig:rho}, following the behavior predicted in Refs.~\cite{Damour:1998jm,Kopeikin:1998ts,Kopeikin:1999ev,Kopeikin:2006gf}.

\begin{figure}[htbp]
    \centering
    \includegraphics[width=\linewidth]{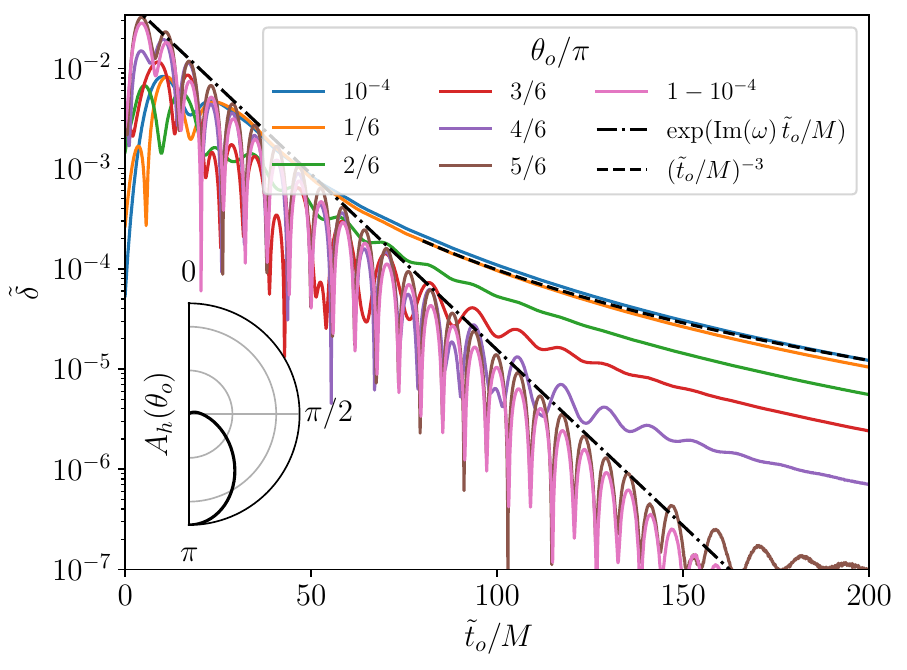}
    \caption{Evolutions of the deflection angles of geodesics terminating at infinity, observed from various inclination angles $\theta_0$ and plotted against observation time $\tilde{t}_o$. The impact parameter is consistently set at $\tilde{\rho} / M = 10$ along the $+\tilde{X}$ axis. The dashed black line indicates that $\tilde{\delta}$ scales proportionally to $\tilde{t}_o^{-3}$.
    {\bf Inset:} 
    The relative strain amplitudes of the GWs, $A_h$, defined as $h \equiv (h_+^2 + h_\times^2)^{1/2} \sim A_h \exp(\mathrm{Im}(\omega) \tilde{t}_o)$, are plotted as a function of the $\theta_o$. The amplitude is most prominent at $\theta_o = \pi$.}
    \label{fig:theta_dep}
\end{figure}

We also present cases with different observer inclination angles $\theta_o$ in Fig.~\ref{fig:theta_dep}, with the impact parameters fixed at $\tilde{\rho} / M = 10$ along the $+\tilde{X}$-axis. Similar to the observations in Fig.~\ref{fig:rho}, the evolutions exhibit two distinct types of behavior: a ringdown-like exponential decay with oscillations, and a subsequent power-law decay proportional to $\tilde{t}_o^{-3}$. The power-law tail becomes more pronounced at smaller $\theta_o$. This effect is due to the polar angle $\theta$ of photon geodesics at $t \approx r_*$ being closer to $\pi$, where the metric perturbation of the $\ell=m=2$ mode is most dominant, as shown in the angular distribution of the GW ringdown in the inset panel. For $\theta_o = \pi(1-10^{-4})$, the deflection angle is predominantly characterized by ringdown behavior, with the oscillation frequency mirroring the ringdown frequency to one part in $10^{10}$.

Furthermore, we consider the effect of incorporating the QNM's mirror mode with equal energy, reflecting a more realistic scenario where the SMBHB was in a quasi-circular orbit before the merger. This setup also exhibits the two previously discussed characteristics: an initial rapid decay during the ringdown phase, followed by a late-time power-law tail, as shown in Supplemental Material~C~\cite{SupplementalMaterial}.

\section{Discussion}

This study explored the dynamic lensing of photons during the ringdown phase after SMBHB mergers. By calculating photon trajectories around a relaxing Kerr BH using backward ray tracing, we demonstrate how metric perturbations during the ringdown significantly influence photon paths. In most cases, the evolutions of photon deflection angles initially exhibit behavior akin to gravitational wave ringdown, providing insights into the fascinating dynamics of spacetime.

The visualization of ringdown through photon lensing opens a new avenue for multi-messenger astronomy. While pulsar timing arrays (PTAs)~\cite{NANOGrav:2023gor,EPTA:2023fyk,Reardon:2023gzh,Xu:2023wog}, astrometry~\cite{Braginsky:1989pv,Book:2010pf,Moore:2017ity}, binary orbits~\cite{Blas:2021mpc,Blas:2021mqw}, and fast radio burst timing~\cite{Lu:2024yuo}
may directly detect ringdown GWs from SMBHB mergers, observations from the Event Horizon Telescope (EHT)~\cite{EventHorizonTelescope:2019dse,EventHorizonTelescope:2022xnr} and future upgrades~\cite{Ayzenberg:2023hfw}, including space missions~\cite{Haworth:2019urs,Johnson:2024ttr,Akiyama:2024msp}, could provide visualizations of a `Waltz' between GW and photon co-propagation. Each pixel on the observer plane could reveal additional insights by dissecting the metric perturbations in the vicinity of the relaxing BH.

A critical question is the feasibility of observing ringdown in electromagnetic channels. This requires the telescope’s Gaussian kernel, represented by $\theta_{\rm res}/2.4$ (where $\theta_{\rm res}$ is the angular resolution), to be finer than the changes in the impact parameter on the observer plane, scaling as $\tilde{\delta} \times r_s / r_o$ with $r_s$ being the distance to the BH. Substituting from Eq.~\eqref{eq:delta_E_rho} and using $E_{\mathrm{tot}}\sim 0.27 M q^2/(1+q)^4$, where $q$ is the mass ratio~\cite{Berti:2007fi}, we derive the requirement:
\begin{equation}
\frac{M}{r_o} > 0.67 \,
\theta_{\mathrm{res}} \beta \left(2 + q + \frac{1}{q}\right),
\end{equation}
where $\beta \equiv \Delta \tilde{\rho}/r_s < 1$ is a geometric factor dependent on the configuration of the light source, the BH, and the observer. The current angular resolution achievable by the EHT is $20\mathrm{\mu as}$, anticipated to improve to $3 \mathrm{\mu as}$ with space-based missions like the Black Hole Explorer (BHEX)~\cite{Johnson:2024ttr, marrone2024blackholeexplorerinstrument, Issaoun_2024, Haworth:2019urs}. Further enhancements in resolution can be achieved by extending the baselines to locations such as the Moon or the second Sun-Earth Lagrange point~\cite{Johnson:2019ljv}.

For a merger involving BHs with comparable mass ratios, assuming $M = 10^{10}\,M_\odot$ as suggested by PTA observations~\cite{NANOGrav:2023hfp}, BHEX could detect lensing signatures from distant sources out to $r_o \sim 12~\mathrm{Mpc}/\beta$. For well-monitored SMBHs such as Sgr A$^*$ and M87$^*$, even mergers with smaller mass ratios become observable, enabling detection of events with $q$ as small as $0.5\beta$. {A source with a small $\beta$—typically located on the far side of the BH, such as a jet feature, an accretion flow near the light ring, or a background star—can therefore significantly enhance the detection prospects.}

In the case of background stars lensed by Sgr A$^*$, Ref.~\cite{Michalowski:2021uxl} predicts that upcoming telescopes like ELT, TMT, and GMT will detect $\mathcal{O}(100)$ stars with angular separations of $\sim$1\,mas and distances of $\sim$1\,pc behind the BH, corresponding to $\beta \sim 10^{-5}$. Given their mas-level angular resolution, these telescopes could detect lensing signatures from intermediate-mass BHs merging into Sgr A$^*$ with mass ratios as small as $q \sim 10^{-3}$.

While our focus here is on ringdown, the underlying formalism applies more broadly to dynamic metric perturbations. For instance, a single SMBH surrounded by a bosonic superradiant cloud could generate sustained periodic perturbations, potentially observable by EHT-class instruments as time-dependent lensing signatures in the accretion flow near the photon ring~\cite{Chen:2022kzv}. Moreover, our approach is applicable to the long-lasting inspiral stage of a binary system, where the accumulation of signal-to-noise ratio could make such searches feasible. A more precise prediction of the apparent position shift of a point source would require generalizing forward ray tracing~\cite{Zhou:2024dbc} in perturbed spacetime.

\begin{acknowledgments}
We thank Maarten van de Meent for helpful discussions and for providing us with useful routines for metric reconstruction. We also extend our thanks to Michael Boyle for his insightful discussions on spin-weighted spherical harmonics.
Z.Z.\ acknowledges financial support from China Scholarship Council (No.~202106040037).
The Center of Gravity is a Center of Excellence funded by the Danish National Research Foundation under grant No. 184. V.C. and Y. C. acknowledge support by VILLUM Foundation (grant no. VIL37766) and the DNRF Chair program (grant no. DNRF162) by the Danish National Research Foundation. V.C. is a Villum Investigator and a DNRF Chair.  V.C. acknowledges financial support provided under the European Union’s H2020 ERC Advanced Grant “Black holes: gravitational engines of discovery” grant agreement no. Gravitas–101052587.
Views and opinions expressed are however those of the author only and do not necessarily reflect those of the European Union or the European Research Council. Neither the European Union nor the granting authority can be held responsible for them. This project has received funding from the European Union's Horizon 2020 research and innovation programme under the Marie Sklodowska-Curie grant agreement No 101007855 and No 101131233. 
Y.C. is supported by the Rosenfeld foundation in the form of an Exchange Travel Grant and by the COST Action COSMIC WISPers CA21106, supported by COST (European Cooperation in Science and Technology).
\end{acknowledgments}

\bibliography{references}

\appendix

\setcounter{equation}{0}
\setcounter{table}{0}
\setcounter{figure}{0}
\setcounter{section}{0}
\setcounter{subsection}{0}

\renewcommand\thesection{\Alph{section}}
\renewcommand\thesubsection{\arabic{subsection}}
\renewcommand\theequation{\thesection.\arabic{equation}}
\renewcommand{\thefigure}{S\arabic{figure}}

\onecolumngrid

\begin{center}
\textbf{\large Supplemental Material: Lensing Tomography of Black Hole Ringdown}
\end{center}

\section{Metric Perturbations During Ringdown}

\subsection{Newman-Penrose formalism} \label{sec:newman_penrose}

For completeness, we review the Newman-Penrose (NP) formalism in this section, adopting the convention from Refs.~\cite{vandeMeent:2016pee, Spiers:2023cip, Pound2020}. Note the overall sign difference in some NP quantities compared to the common convention, as seen, for example, in Ref.~\cite{Teukolsky:1973ha}. This ensures that the background values remain consistent with cases where the metric signature is $(+,-,-,-)$.
In the NP formalism, four null vectors $e_{(\mu)}^a$ are defined as
\begin{equation}
e_{(\mu)}^a = \left\{e_{(0)}^a, e_{(1)}^a, e_{(2)}^a, e_{(3)}^a\right\} \equiv \left\{l^a, n^a, m^a, \bar{m}^a\right\}
\end{equation}
where $l^a$ and $n^a$ are real null vectors, and $m^a$ and $\bar{m}^a$ are complex null vectors. They satisfy
\begin{equation}
l^a n_a=-1 \quad \text{and}\quad m^a \bar{m}_a=1 \,.
\end{equation}
The metric can be obtained via
\begin{equation}
g_{a b} = -2 l_{(a} n_{b)}+2 m_{(a} \bar{m}_{b)} \, .
\end{equation}
The directional covariant derivatives are defined as
\begin{equation}
\hat{D} \equiv l^a \nabla_a, \,\hat{\Delta} \equiv n^a \nabla_a, \,\hat{\delta} \equiv m^a \nabla_a,\,  \hat{\bar{\delta}} \equiv \bar{m}^a \nabla_a .
\end{equation}
The Ricci rotation coefficients are given by
\begin{equation}
\gamma_{(a)(b)(c)} \equiv g_{\mu \lambda} e_{(a)}^\mu e_{(c)}^\nu \nabla_\nu e_{(b)}^\lambda \,,
\end{equation}
which can be explicitly written as
\begin{equation}
\begin{aligned}
\kappa &\equiv-\gamma_{(3)(1)(1)}\,,\,\varpi \equiv-\gamma_{(2)(4)(1)}\,,\,\epsilon \equiv-\frac{1}{2} \left(\gamma_{(2)(1)(1)}+\gamma_{(3)(4)(1)}\right), \\
\tau &\equiv-\gamma_{(3)(1)(2)}\,,\,\nu \equiv-\gamma_{(2)(4)(2)}\,,\,\gamma \equiv-\frac{1}{2}\left(\gamma_{(2)(1)(2)}+\gamma_{(3)(4)(2)}\right), \\
\sigma &\equiv-\gamma_{(3)(1)(3)}\,,\,\mu \equiv-\gamma_{(2)(4)(3)}\,,\,\beta \equiv-\frac{1}{2} \left(\gamma_{(2)(1)(3)}+\gamma_{(3)(4)(3)}\right), \\
\rho &\equiv-\gamma_{(3)(1)(4)}\,,\,\lambda \equiv-\gamma_{(2)(4)(4)}\,,\,\alpha \equiv-\frac{1}{2}\left(\gamma_{(2)(1)(4)}+\gamma_{(3)(4)(4)}\right) \,.
\end{aligned}
\end{equation}
The formal adjoint $\mathcal{L}^\dag$ of an operator
\begin{equation}
\mathcal{L}=e_{(\mu)}^a \nabla_a \,
\end{equation}
can be written as \cite{Wald:1978vm, Dias:2009ex, Cartas-Fuentevilla:1997ije}
\begin{equation}
\mathcal{L}^\dag =-\mathcal{L}-\nabla_a e_{(\mu)}^a
\end{equation}
Therefore, the formal adjoint directional derivative operators are
\begin{equation}
\begin{aligned}
\hat{D}^{\dagger} &=-(\hat{D}+\varepsilon+\bar{\varepsilon}-\rho-\bar{\rho})\,,\,
\hat{\Delta}^{\dagger} =-(\hat{\Delta}-\gamma-\bar{\gamma}+\mu+\bar{\mu}), \\
\hat{\delta}^{\dagger} &=-(\hat{\delta}+\beta-\bar{\alpha}-\tau+\bar{\varpi})\,,\,
\hat{\bar{\delta}}^{\dagger} =-(\hat{\bar{\delta}}+\bar{\beta}-\alpha-\bar{\tau}+\varpi) \,.
\end{aligned}
\end{equation}
In the Kerr background, we use the Kinnersley tetrad as follows:
\begin{equation}
\begin{aligned}
e^\mu_{\mathrm{(1)}} = l^\mu & =\frac{1}{\Delta_{\mathrm{BL}}}\left(r^2+a^2, \Delta_{\mathrm{BL}}, 0, a\right)\,,\quad e^\mu_{\mathrm{(2)}} =n^\mu  =\frac{1}{2 \Sigma_{\mathrm{BL}}}\left(r^2+a^2,-\Delta_{\mathrm{BL}}, 0, a\right), \\
e^\mu_{\mathrm{(3)}} =m^\mu & =-\frac{\bar{\rho}}{\sqrt{2}}\left(i a \sin\theta, 0, 1, \frac{i}{\sin\theta}\right)\,,\quad
e^\mu_{\mathrm{(4)}} =\bar{m}^\mu  =\frac{\rho}{\sqrt{2}}\left(i a \sin\theta, 0, -1, \frac{i}{\sin\theta}\right),
\end{aligned}
\end{equation}
where $\rho = - 1 / (r - i a \cos\theta)$ and the overbar denotes complex conjugation.  $\Delta_{\mathrm{BL}}$ and $\Sigma_{\mathrm{BL}}$ are defined as
\begin{equation} 
\Delta_{\mathrm{BL}}\equiv r^{2}-2 M r+a^{2}\,, \quad \Sigma_{\mathrm{BL}}\equiv r^{2}+a^{2} \cos^{2} \theta\,.
\end{equation}
The Ricci rotation coefficients can be written explicitly as
\begin{equation}
\kappa=\lambda=\nu=\sigma=\epsilon=0 \,,
\end{equation}
and
\begin{equation}
\begin{aligned}
& \rho=- 1 /(r-i a \cos \theta) \,,\, \beta=-\bar{\rho} \cot \theta /(2 \sqrt{2}) \,,\, \varpi=i a \rho^2 \sin \theta / \sqrt{2} \,, \\
& \tau=-i a \rho \bar{\rho} \sin \theta / \sqrt{2} \,,\,\mu=\rho^2 \bar{\rho} \Delta / 2 \,,\, \gamma=\mu+\rho \bar{\rho} (r-M) / 2 \,,\, \alpha=\varpi - \bar{\beta} \,.
\end{aligned}
\end{equation}
The Weyl scalars are defined as
\begin{equation}
\begin{aligned}
& \Psi_0 \equiv C_{(1)(3)(1)(3)} = C_{\mu \nu \rho \sigma} l^\mu m^\nu l^\rho m^\sigma \,,\, \Psi_1 \equiv C_{(1)(2)(1)(3)} = C_{\mu \nu \rho \sigma} l^\mu n^\nu l^\rho m^\sigma \,, \\
& \Psi_2 \equiv C_{(1)(3)(4)(2)} = C_{\mu \nu \rho \sigma} l^\mu m^\nu \bar{m}^\rho n^\sigma \,,\, \Psi_3 \equiv C_{(1)(2)(4)(2)} = C_{\mu \nu \rho \sigma} l^\mu n^\nu \bar{m}^\rho n^\sigma \,, \\
& \Psi_4 \equiv C_{(2)(4)(2)(4)} = C_{\mu \nu \rho \sigma} n^\mu \bar{m}^\nu n^\rho \bar{m}^\sigma \,,
\end{aligned}
\end{equation}
where $C_{\mu \nu \rho \sigma}$ is the Weyl tensor. In the Kerr background, we have $\Psi_0 = \Psi_1 = \Psi_3 = \Psi_4 = 0$ and
\begin{equation}
\Psi_2 = \rho^3 M \,.
\end{equation}
Other useful equations derived from the Bianchi identities are
\begin{equation}
\hat{D} \Psi_2 = 3 \rho \Psi_2 \,,\, 
\hat{\Delta} \Psi_2 = -3 \mu \Psi_2 \,,\,\hat{\delta} \Psi_2 = 3 \tau \Psi_2 \,,\,\hat{\bar{\delta}} \Psi_2 = -3 \varpi \Psi_2 \,. 
\end{equation}

\subsection{Metric Reconstruction} \label{sec:metric_reconstruction}

Cohen, Chrzanowski, and Kegeles (CCK) formulated a method for reconstructing vacuum metric perturbations within a radiation gauge \cite{Chrzanowski:1975wv, Cohen:1974cm, Kegeles:1979an}. Here, we briefly review the procedure of CCK reconstruction, which is detailed in Ref.~\cite{Toomani:2021jlo, Spiers:2023cip, Green:2019nam, Campanelli:1998jv, Whiting:2005hr, Stewart:1978tm}. Many of our calculations heavily depend on the Newman-Penrose (NP) formalism, adhering to the conventions outlined in Appendix~\ref{sec:newman_penrose}.

We begin by introducing four operators: $\hat{\mathcal{E}}$, $\hat{\mathcal{O}}$, $\hat{\mathcal{S}}$, and $\hat{\mathcal{T}}$, each with their respective formal adjoints:
\begin{itemize}
    \item The linearized Einstein operator $\hat{\mathcal{E}}$, which is defined as
\begin{equation}
\begin{split}
\hat{\mathcal{E}}_{ab}(h) =\frac{1}{2}\left[-\nabla^c \nabla_c h_{a b}-\nabla_a \nabla_b h_c{}^c+2 \nabla^c \nabla_{(a} h_{b) c}  +g_{a b}\left(\nabla^c \nabla_c h_d{}^d-\nabla^c \nabla^d h_{c d}\right)\right] \,.
\end{split}
\end{equation}
The operator $\hat{\mathcal{E}}$ is self-adjoint, as indicated by $\hat{\mathcal{E}} = \hat{\mathcal{E}}^\dag$, and it maps a metric perturbation to its corresponding linearized Einstein tensor.
\item The Teukolsky operator $\hat{\mathcal{O}}$, which can be written as \cite{Teukolsky:1973ha}
\begin{equation}
\begin{aligned}
\hat{\mathcal{O}}_0 = & \left(\hat{D}-3 \epsilon+ \bar{\epsilon}-4 \rho-\bar{\rho}\right) \left(\hat{\Delta}-4 \gamma+\mu\right) - \left(\hat{\delta}+ \bar{\varpi}- \bar{\alpha}-3 \beta-4 \tau\right)\left(\hat{\bar{\delta}}+\varpi-4 \alpha\right) -3 \Psi_2 \,, \\
\hat{\mathcal{O}}_4 = & \left(\hat{\Delta}+3 \gamma-\bar{\gamma}+4 \mu+\bar{\mu} \right) \left(\hat{D}+4 \epsilon-\rho \right)  - \left(\hat{\overline{\delta}}-\bar{\tau}+\bar{\beta}+3 \alpha+4 \varpi \right) \left( \hat{\delta}-\tau+4 \beta \right)-3 \Psi_2 \,.
\end{aligned}
\end{equation}
Then the Teukolsky equation with spin $\pm 2$ can be expressed as
\begin{equation}
\hat{\mathcal{O}}_0 \Psi_0^{(1)} = 8 \pi T_0 \,, \,
\hat{\mathcal{O}}_4 \Psi_4^{(1)} = 8 \pi T_4 \,.
\end{equation}
Their separable form, the Teukolsky master equation, can be written as
\begin{equation}
\hat{\mathcal{O}}_s \psi_s = 8 \pi \Sigma T_s \,, \label{eq:teukolsky_master_equation}
\end{equation}
where the spin $s$ is designated as either $\pm 2$. For $s = -2$,
\begin{equation}
\psi_{-2} = \rho^{-4} \Psi_4^{(1)} \,, \hat{\mathcal{O}}_{-2} = 2 \Sigma \rho^{-4} \hat{\mathcal{O}}_4 \rho^{4} \,, T_{-2} = 2 \rho^{-4} T_4 \,;
\end{equation}
conversely, when $s = +2$,
\begin{equation}
\psi_{+2} = \Psi_0^{(1)}, \, \hat{\mathcal{O}}_{+2} =  2 \Sigma \hat{\mathcal{O}}_0   \,, T_{+2} = 2 T_0 \, .
\end{equation}
\item The operator $\hat{\mathcal{S}}$, which acts on an energy-momentum tensor $T_{ab}$ to yield the source $T_4$ of the Teukolsky equation, (Eq.~(2.15) in Ref.~\cite{Teukolsky:1973ha})
\begin{equation}
\begin{aligned}
\hat{\mathcal{S}}^{a b} = & \left( \hat{\Delta}+3 \gamma-\bar{\gamma}+4 \mu+\bar{\mu} \right) \left[(\hat{\overline{\delta}}-2 \bar{\tau}+2 \alpha) n^{(a} \bar{m}^{b)}-(\hat{\Delta}+2 \gamma-2 \bar{\gamma}+\bar{\mu}) \bar{m}^a \bar{m}^b\right] \\
& + \left(\hat{\overline{\delta}}-\bar{\tau}+\bar{\beta}+3 \alpha+4 \varpi \right) \left[ \left(\hat{\Delta}+2 \gamma+2 \bar{\mu} \right) n^{(a} \bar{m}^{b)}- \left( \hat{\overline{\delta}}-\bar{\tau}+2 \bar{\beta}+2 \alpha \right) n^a n^b\right],
\end{aligned}
\end{equation}
\item The operator $\hat{\mathcal{T}}$, which acts on a metric perturbation $h_{ab}$ to yield the perturbed Weyl scalar $\Psi_4^{(1)}$.
\begin{equation}
\begin{aligned}
\hat{\mathcal{T}}^{a b}= & -\frac{1}{2}\left\{ \left(\hat{\overline{\delta}}-\bar{\tau}+3 \alpha+\bar{\beta} \right) \left(\hat{\overline{\delta}}-\bar{\tau}+2 \alpha+2 \bar{\beta} \right) n^a n^b\right. \\
& + \left(\hat{\Delta}+\bar{\mu}+3 \gamma-\bar{\gamma} \right)\left(\hat{\Delta}+\bar{\mu}+2 \gamma-2 \bar{\gamma} \right) \bar{m}^a \bar{m}^b \\
& -\left[ \left(\hat{\Delta}+\bar{\mu}+3 \gamma-\bar{\gamma} \right) \right. \left(\hat{\overline{\delta}}-2 \bar{\tau}+2 \alpha \right) \\
& \left.\left.+ \left(\hat{\overline{\delta}}-\bar{\tau}+3 \alpha+\bar{\beta} \right) \left(\hat{\Delta}+2 \bar{\mu}+2 \gamma \right) \right] n^{(a} \bar{m}^{b)}\right\}
\end{aligned}
\end{equation}
\end{itemize}
\begin{figure}[h!]
    \centering
    \schemestart
      $h_{ab}$
      \arrow{<=>[$\hat{\mathcal{E}}$][$\hat{\mathcal{E}}^\dag$]} $T_{ab}$
      \arrow{<=>[$\hat{\mathcal{S}}$][$\hat{\mathcal{S}}^\dag$]}[-90] $T_4$
      \arrow{<=>[$\hat{\mathcal{O}}_4$][$\hat{\mathcal{O}}^\dag_4$]}[-180] $\Psi_4^{(1)}$
      \arrow{<=>[$\hat{\mathcal{T}}^\dag$][$\hat{\mathcal{T}}$]}[90]
    \schemestop
    \caption{Relation of operators for $s = -2$.}
    \label{fig:metric_reconstruction}
\end{figure}

As pointed out by Wald \cite{Wald:1978vm}, one can find the following operator identity
\begin{equation}
\hat{\mathcal{O}}_4 \hat{\mathcal{T}} = \hat{\mathcal{S}} \hat{\mathcal{E}}\,,
\end{equation}
and its formal adjoint, $\hat{\mathcal{T}}^{\dag} \hat{\mathcal{O}}^{\dag}_4 = \hat{\mathcal{E}} \hat{\mathcal{S}}^{\dag}$, is also valid.
Then, assuming a complex potential $\Phi$, also known as Herz potential, satisfies the adjoint Teukolsky equation $\hat{\mathcal{O}}^\dag_4 \Phi = 0$, we can consequently have
\begin{equation}
\hat{\mathcal{E}}_{ab}(h) = 2 \operatorname{Re}(\hat{\mathcal{T}}^\dag_{ab} \hat{\mathcal{O}}^\dag_4 \Phi) = 0
\end{equation}
Therefore, under the assumption, $h_{ab}$ provides a solution to the linearized Einstein equation, and can be expressed as
\begin{equation}
h_{ab}=2 \operatorname{Re}\left(\hat{\mathcal{S}}_{a b}^{\dagger} \Phi\right)  \,.
\end{equation}

Chrzanowski introduced two radiation gauges: the \textit{ingoing radiation gauge} (IRG) and the \textit{outgoing radiation gauge} (ORG) \cite{Chrzanowski:1975wv}. In this paper, we choose ORG, which can be written as
\begin{equation}
h_{ab} n^b = 0 = g^{ab}h_{ab} \,.
\end{equation}
Under this gauge condition, the metric satisfies
\begin{equation}
h_{n n}=0=h_{l n}=0=h_{n m}=0=h_{n \bar{m}}=0=h_{m \bar{m}} \,.
\end{equation}
The Hertz potential $\Phi_{\mathrm{ORG}}$ fulfills the adjoint Teukolsky equation $\mathcal{O}^\dag_4 \Phi_{\mathrm{ORG}} = 0$, where
\begin{equation}
\mathcal{O}^\dag_4 = \left(\hat{D}-3 \epsilon+\bar{\epsilon}-\bar{\rho} \right) \left(\hat{\Delta}-4 \gamma-3 \mu \right)- \left(\hat{\delta}-3 \beta-\bar{\alpha}+\bar{\varpi} \right)\left(\hat{\bar{\delta}}-4 \alpha-3 \varpi \right)-3 \Psi_2 \,.
\end{equation}
One can verify $\mathcal{O}^\dag_4 \Phi_{\mathrm{ORG}} = 0$ is equivalent to the homogeneous Teukolsky equation for a field with $s = +2$, that is,
\begin{equation}
\hat{\mathcal{O}}_4^\dag \Phi_{\mathrm{ORG}}=\rho^{-4} \mathcal{O}_0\left(\rho^{4} \Phi_{\mathrm{ORG}}\right) = 0 \,,
\end{equation}
and the metric component can be written as \cite{Dias:2009ex}
\begin{equation}
\begin{aligned}
h_{a b}= &  \left\{n_{(a} \bar{m}_{b)} \left[ \left(\hat{\bar{\delta}}+\bar{\beta}-3 \alpha+\bar{\tau}+\varpi \right) \left(\hat{\Delta}-4 \gamma-3 \mu \right)\right.\right. \\
& \left. + \left(\hat{\Delta}-3 \gamma-\bar{\gamma}+\mu-\bar{\mu} \right)\left(\hat{\bar{\delta}}-4 \alpha-3 \varpi \right) \right] \\
& -n_a n_b \left(\hat{\bar{\delta}}-\bar{\beta}-3 \alpha+\varpi \right)\left(\hat{\bar{\delta}}-4 \alpha-3 \varpi \right)  \\
& \left. -\bar{m}_a \bar{m}_b \left(\hat{\Delta}-3 \gamma+\bar{\gamma}+\mu \right) \left(\hat{\Delta}-4 \gamma-3 \mu \right) \right\} \Phi_{\mathrm{ORG}} + \text{c.c.} \,,
\end{aligned}
\end{equation}
where ``c.c.'' stands for the complex conjugate part of the whole object. If we define $\mathring{\Phi}_{\mathrm{ORG}} = \rho^4 \Phi_{\mathrm{ORG}}$, then the non-zero metric component can be written as \cite{Lousto:2002em, vandeMeent:2015lxa}
\begin{equation}
\begin{aligned}
& h_{ll} \equiv h_{\mu \nu} l^\mu l^\nu = \hat{\mathcal{H}}_{ll}^\dag \mathring{\Phi}_{\mathrm{ORG}} + \text{c.c.}\,,\, h_{lm} \equiv h_{\mu \nu} l^\mu m^\nu = \hat{\mathcal{H}}_{lm}^\dag \mathring{\Phi}_{\mathrm{ORG}}, \\
& h_{l\bar{m}} \equiv \bar{h}_{lm}\,,\, h_{mm} \equiv h_{\mu \nu} m^\mu m^\nu = \hat{\mathcal{H}}_{mm}^\dag \mathring{\Phi}_{\mathrm{ORG}},
\end{aligned}
\end{equation}
where
\begin{equation}
\begin{aligned}
\hat{\mathcal{H}}_{ll}^\dag &= - \rho^{-4} \left(\hat{\bar{\delta}}-3 \alpha-\bar{\beta}+5 \varpi \right) \left(\hat{\bar{\delta}}-4 \alpha+\varpi \right) \,, \\
\hat{\mathcal{H}}_{lm}^\dag &= - \frac{\rho^{-4}}{2} \left\{ \left(\hat{\bar{\delta}}-3 \alpha+\bar{\beta}+5 \varpi+\bar{\tau} \right)\left(\hat{\Delta}+\mu-4 \gamma \right) \right. \\
& \left. + \left(\hat{\Delta}+5 \mu-\bar{\mu}-3 \gamma-\bar{\gamma} \right) \left(\hat{\bar{\delta}}-4 \alpha+\varpi \right) \right\} \,, \\
\hat{\mathcal{H}}_{mm}^\dag &= - \rho^{-4} \left(\hat{\Delta}+5 \mu-3 \gamma+\bar{\gamma}\right) \left(\hat{\Delta}+\mu-4 \gamma \right) \,.
\end{aligned}
\end{equation}
While $\mathring{\Phi}_{\mathrm{ORG}}$ does indeed satisfy the vacuum Teukolsky equation with $s = +2$, the field does not correspond to either the Weyl scalar $\Psi^{\mathrm{(1)}}_0$ or $\Psi^{\mathrm{(1)}}_4$ originating from the reconstructed metric $h_{ab}$. From Fig.~\ref{fig:metric_reconstruction}, it is clear that $\Psi^{\mathrm{(1)}}_4$ can be obtained via a fourth-order ordinary differential equation (ODE) along ingoing null rays \cite{Ori:2002uv, vandeMeent:2015lxa, Spiers:2023cip}
\begin{equation}
\Psi_{4}^{(1)} = \hat{\mathcal{T}}^{ab} h_{a b} \,. %
\end{equation}
We can reduce this circularity condition to what is known as a radial inversion relation \cite{Spiers:2023cip,Kofron:2020mqw, Aksteiner:2016mol, Deadman_2011, Pound2020}
\begin{equation}
\Psi_{4}^{(1)} = \frac{1}{2} \left(\hat{\Delta}+3 \gamma-\bar{\gamma} \right) \left(\hat{\Delta}+2 \gamma-2 \bar{\gamma} \right) \left(\hat{\Delta}+\gamma-3 \bar{\gamma} \right) \left( \hat{\Delta}-4 \bar{\gamma} \right) \bar{\Phi}_{\mathrm{ORG}} \,, \label{eq:fourth_order_NP}
\end{equation}
which can be solved in vacuum algebraically by applying the separation of variables in the next section.

\subsubsection{QNM to Hertz potential}

In Boyer-Lindquist (BL) coordinates, the Teukolsky operator $\hat{\mathcal{O}}_s$ with spin $s$ in Eq.~\eqref{eq:teukolsky_master_equation} can be written as~\cite{Teukolsky:1973ha}
\begin{align}
\hat{\mathcal{O}}_s \psi_s &= {\left[\frac{\left(r^2+a^2\right)^2}{\Delta_{\mathrm{BL}}}-a^2 \sin ^2 \theta\right] \frac{\partial^2{ } \psi_s}{\partial t^2}} 
+\frac{4 M a r}{\Delta_{\mathrm{BL}}} \frac{\partial^2 \psi_s}{\partial t \partial \phi}+\left[\frac{a^2}{\Delta}-\frac{1}{\sin ^2 \theta}\right] \frac{\partial^2 \psi_s}{\partial \phi^2}\nonumber \\
&-\Delta_{\mathrm{BL}}^{-s} \frac{\partial}{\partial r}\left(\Delta_{\mathrm{BL}}^{s+1} \frac{\partial \psi_s}{\partial r}\right)-\frac{1}{\sin \theta} \frac{\partial}{\partial \theta}\left(\sin \theta \frac{\partial \psi_s}{\partial \theta}\right) -2 s\left[\frac{a(r-M)}{\Delta_{\mathrm{BL}}}+\frac{i \cos \theta}{\sin ^2 \theta}\right] \frac{\partial \psi_s}{\partial \phi} \nonumber\\
&-2 s\left[\frac{M\left(r^2-a^2\right)}{\Delta_{\mathrm{BL}}}-r-i a \cos \theta\right] \frac{\partial \psi_s}{\partial t} +\left(s^2 \cot ^2 \theta-s\right) \psi_s
\end{align}
where $\psi_s$ is a field with spin $s$.
In the vacuum case $T_s = 0$, we can apply the separation of variables by letting
\begin{equation}
\psi_s(t, r, \theta, \phi) = \frac{1}{\sqrt{2\pi}} \sum_{l,m,n} 
{}_s Z_{\ell m n} \, {}_s S_{\ell m n}(\theta) \, {}_{s}R_{\ell m n}(r) e^{-i \omega_{m n} t + i m \phi} \,, \label{eq:psi_s_mode}
\end{equation}
With $z = \cos\theta$, ${}_{s} S_{\ell m n}(\theta)$ satisfy
\begin{equation}
\partial_z\left[\left(1-z^2\right) \partial_z\left[{}_{s}S_{\ell m n}(\theta)\right] \right]
+\left[(c z)^2-2 c s z+s + A^{\ell m n}_s(c)
-\frac{(m+s z)^2}{1-z^2}\right] {}_{s}S_{\ell m n}(\theta)=0
\end{equation}
where $c = a \omega$, and ${}_{s}S_{\ell m n}(\theta) e^{im\phi}$ represents the spin-weighted spheroidal harmonics, with their phase is consistent with Ref.~\cite{Cook:2014cta}.
To normalize the spin-weighted spheroidal harmonics, we apply the following normalization condition
\begin{equation}
\int_0^{\pi} {}_s S_{\ell m n}(\theta) \bar{S}_{\ell m n}(\theta) \sin\theta \dd{\theta} = 1 \,.\label{eq:S_normalization}
\end{equation}
The radial function $R(r)$ is required to fulfill
\begin{equation}
\Delta^{-s}_{\mathrm{BL}} \frac{d}{d r}  {\left[\Delta^{s+1}_{\mathrm{BL}} \frac{\dd{ {}_s R_{\ell m n}(r)}}{\dd{r}}\right] }  \quad +\left[\frac{K^2-2 i s(r-M) K}{\Delta_{\mathrm{BL}}}+4 i s \omega r-\lambdabar_{\ell m n}\right] {}_s R_{\ell m n}(r)=0 \,. \label{eq:radial_teukolsky}
\end{equation}
where
\begin{equation}
K_{mn}  \equiv\left(r^2+a^2\right) \omega_{mn} - a m \,,\,
{}_s \lambdabar_{\ell m n}  \equiv {}_s A_{\ell m n} + a^2 \omega_{m n}^2-2 a m \omega_{m n} .
\end{equation}
Here ${ }_s A_{\ell m n}$ is the angular separation constant.

In the following, we will revisit the procedure for obtaining the Hertz potential corresponding to specific Weyl scalar $\Psi^{\mathrm{(1)}}_0$ or $\Psi^{\mathrm{(1)}}_4$, which is discussed in Ref.~\cite{Ori:2002uv, vandeMeent:2015lxa}. According to the previously mentioned condition, $\hat{\mathcal{O}}_{+2} \mathring{\Phi}_{\mathrm{ORG}} = 0$, we can expand $\mathring{\Phi}_{\mathrm{ORG}}$ in modes
\begin{equation}
\mathring{\Phi}_{\mathrm{ORG}}= \frac{1}{\sqrt{2\pi}} \sum_{\ell, m, n} C_{\ell m n} \, {}_{+2}R_{\ell m n}(r) \, {}_{+2}S_{\ell m n}(\theta) e^{i m \phi-i \omega_{mn} t} \,, \label{eq:hertz_mode_basis}
\end{equation}
where $C_{\ell m n}$ are constant coefficients. In addition, the Hertz potential $\mathring{\Phi}_{\mathrm{ORG}}$ also satisfies Eq.~\eqref{eq:fourth_order_NP}, which can be simplified using the following relationship
\begin{equation}
\left( \hat{\Delta} - m \bar{\gamma} + n \gamma \right) f = \Delta_{\mathrm{BL}}^{(n-m)/2} \bar{\rho}^{-m} \rho^{n} \hat{\Delta} \left[\Delta_{\mathrm{BL}}^{(m-n)/2} \bar{\rho}^{m} \rho^{-n} f\right] \,.
\end{equation}
In this equation, $f = f(t, r, \theta, \phi)$ can represent any function. Then Eq.~\eqref{eq:fourth_order_NP} can be written as \cite{Lousto:2002em, vandeMeent:2015lxa, Ori:2002uv, Pound2020}
\begin{equation}
\frac{1}{32} \Delta_{\mathrm{BL}}^2\left(\hat{\mathcal{D}}_0^{\dagger}\right)^4 \left[ \Delta_{\mathrm{BL}}^2 \bar{\mathring{\Phi}}_{\mathrm{ORG}} \right] = \rho^{-4} \Psi_4^{\mathrm{(1)}} \,. \label{eq:fourth_order_condition}
\end{equation}
Here, the operator $\hat{\mathcal{D}}_0^{\dagger}$ is defined as:
\begin{equation}
\hat{\mathcal{D}}_0^{\dagger}=\partial_r-\frac{\left(r^2+a^2\right) \partial_t+a \partial_\phi}{\Delta_{\mathrm{BL}}} \,. 
\end{equation}
Note that Eq.~\eqref{eq:fourth_order_condition} can be solved easily on the mode basis given by Eq.~\eqref{eq:hertz_mode_basis}, using the operator
\begin{equation}
\hat{\mathcal{D}}_{m n}^{\dagger}= \partial_r+i \frac{K_{mn}}{\Delta_{\mathrm{BL}}}= \partial_r+i \frac{\omega_{m n}\left(r^2+a^2\right)-m a}{\Delta_{\mathrm{BL}}}  \,,
\end{equation} %
which acts on radial modes.
The coefficient can be fixed by considering the asymptotic limit approaching infinity and the horizon. This analysis was previously carried out in Ref.~\cite{Ori:2002uv, vandeMeent:2015lxa} for real $\omega$. Due to our choice of gauge (ORG), the results are similar to those in Section III.C of Ref.~\cite{vandeMeent:2015lxa}. Here, we briefly review and extend the procedure to complex $\omega$.
\begin{equation}
\frac{1}{32} \Delta_{\mathrm{BL}}^2\left(\hat{\mathcal{D}}_0^{\dagger}\right)^4 \left[ \Delta_{\mathrm{BL}}^2 \bar{\mathring{\Phi}}_{\mathrm{ORG}} \right] = \sum_{\ell,m,n} \frac{\bar{C}_{\ell m n}}{32 \sqrt{2\pi}} \Delta_{\mathrm{BL}}^2 \left(\partial_r-i \frac{\bar{K}_{mn}}{\Delta_{\mathrm{BL}}}\right)^4 \left[ \Delta_{\mathrm{BL}}^2 \, {}_{+2}\bar{R}_{\ell m n}(r) \, {}_{+2}S_{\ell m n}(\theta) e^{-i m \phi + i \bar{\omega}_{mn} t}  \right]
\end{equation}
By relabelling $(m, n)$ to $(-m,-n)$ and employing the identities $ {}_s S_{\ell m n}(\theta)=(-1)^{s+m} {}_{-s} S_{l-m-n}(\theta)$, ${}_s \bar{R}_{\ell m n} = {}_s R_{l-m-n}$, $\bar{K}_{mn} = -K_{-m -ns}$ and $\bar{\omega}_{mn} = -\omega_{-m -n}$, the equation transforms into
\begin{equation}
\frac{1}{32} \Delta_{\mathrm{BL}}^2\left(\hat{\mathcal{D}}_0^{\dagger}\right)^4 \left[ \Delta_{\mathrm{BL}}^2 \bar{\mathring{\Phi}}_{\mathrm{ORG}} \right] = \sum_{\ell,m,n} (-1)^m \frac{\bar{C}_{\ell -m -n}}{32 \sqrt{2\pi}} \Delta_{\mathrm{BL}}^2 
\left(\hat{\mathcal{D}}_{mn}^{\dagger}\right)^4\left[ \Delta_{\mathrm{BL}}^2 \, {}_{+2}\bar{R}_{\ell m n}(r) \, {}_{-2} S_{\ell m n}(\theta) e^{i m \phi - i \omega_{mn} t}  \right]
\end{equation}
When considering Eq.~\eqref{eq:psi_s_mode}, we obtain
\begin{equation}
{}_{-2} Z_{\ell m n} \, {}_{-2} R_{\ell m n} (r) = \frac{(-1)^m}{32} \bar{C}_{\ell -m -n} \Delta_{\mathrm{BL}}^2 \left(\hat{\mathcal{D}}_{mn}^{\dagger}\right)^4 \left[ \Delta_{\mathrm{BL}}^2 \, {}_{+2} R_{\ell m n}(r) \right] \,.
\end{equation}
The asymptotic behavior of ${}_s R_{\ell m n}$ is as follows \cite{Teukolsky:1974yv, Cook:2014cta}:
\begin{equation}
\lim _{r^* \rightarrow \infty}  {}_s R_{\ell m n}(r^*) \sim \begin{cases}\frac{e^{-i \omega r^*}}{r} & : \text { ingoing wave } \\ \frac{e^{i \omega r^*}}{r^{2 s+1}} & : \text { outgoing wave. }\end{cases}
\end{equation}
\begin{equation}
\lim _{r^* \rightarrow-\infty} {}_s R_{\ell m n} \left(r^*\right) \sim \begin{cases}e^{i k_{mn} r^*} & : \text { ingoing (out of } \mathrm{BH}) \\ \Delta_{\mathrm{BL}}^{-s} e^{-i k_{mn} r^*} & : \text { outgoing (into BH) }\end{cases}
\end{equation}
where $k_{m n}=\omega_{m n}-\frac{m a}{2 M r_{+}}$, and $r^*$ is the tortoise coordinate which is given by
\begin{equation}
r^*=r+\frac{2 M}{r_{+}-r_{-}}\left[r_{+} \ln \left(\frac{r-r_{+}}{2M}\right)-r_{-} \ln \left(\frac{r-r_{-}}{2M}\right)\right]
\end{equation}

If $f(r)$ is a solution to the radial Teukolsky equation Eq.~\eqref{eq:radial_teukolsky} with spin $s$, then $\Delta_{\mathrm{BL}}^s \bar{f}(r)$ is a solution to the radial Teukolsky equation with spin $-s$. For generic modes with $\omega_{m n} \neq 0$, the asymptotic behavior suggests that this relation swaps physical (retarded) boundary conditions with unphysical (advanced) ones. We denoted the relationship as
\begin{equation}
{ }_{-s} R^\#_{\ell m n} \equiv \Delta_{\mathrm{BL}}^s \, { }_s \bar{R}_{\ell m n}.
\end{equation}
With this in mind, we have
\begin{equation}
{}_{-2} Z_{\ell m n} \, {}_{-2} R_{\ell m n} (r) = \frac{(-1)^m}{32} \bar{C}_{\ell -m -n} \Delta_{\mathrm{BL}}^2 \left(\hat{\mathcal{D}}_{mn}^{\dagger}\right)^4 \left[ { }_{-2} R^\#_{\ell m n} \right] \,. \label{eq:ZC_relation}
\end{equation}
Inserting the asymptotic behavior into Eq.~\eqref{eq:ZC_relation}, we find
\begin{equation}
C_{\ell m n}= \frac{2}{\omega_{m n}^4} (-1)^{\ell+m}  {}_{-2} Z_{\ell m n}
\end{equation}
Here, we used the following relationship
\begin{equation}
{}_{-2}\bar{Z}_{l m n}=(-1)^l {}_{-2}Z_{l-m-n} \,.
\end{equation}
Consequently, once we have the solution set $\{\omega, \Psi_4^{\mathrm{(1)}}\}$ for the Teukolsky equation with $s = -2$, we are able to reconstruct the first-order metric perturbation $h_{ab}$.
After reconstructing the metric, we verify its validity by substituting it into the linearized Einstein equation $\hat{\mathcal{E}}_{ab}(h) = 0$.

We calculate the quasinormal mode using the \texttt{qnm} package \cite{Stein:2019mop}, which employs Leaver's method to the radial equation and the Cook-Zalutskiy spectral approach to the angular sector \cite{leaver1986solutions, Cook:2014cta}. Furthermore, we use the pseudospectral method to get solutions to the radial Teukolsky equation. Here we briefly illustrate our steps. We convert the asymptotic behavior into coordinate $r$ \cite{Cook:2014cta}
\begin{equation}
\lim _{r \rightarrow \infty} {}_s R_{\ell m n}(r) \sim r^{-1-2s+2 i \omega_{mn} M}e^{i \omega_{mn} r}
\end{equation}
\begin{equation}
\lim _{r \rightarrow r_{+}} {}_s R_{\ell m n} (r) \sim \left(r-r_{+}\right)^{-s-i \sigma_{+}}
\end{equation}
where $\sigma_+ = \frac{2 \omega_{mn} M r_{+}-m a}{r_{+}-r_{-}}$.
Then we can define a new radial function ${}_s \tilde{R}_{\ell m n} (r)$ by
\begin{equation}
{}_s R_{\ell m n} (r) = \left(r-r_{+}\right)^{-s-i \sigma_{+}} r^{-1-s+2 i \omega_{mn} M+i\sigma_{+}}  e^{i \omega_{mn} r} {}_s \tilde{R}_{\ell m n} (r) \,,
\end{equation}
where ${}_s \tilde{R}_{\ell m n} (r)$ is regular at BH horizon and infinity, and we normalize it by applying
\begin{equation}
\lim_{r\to\infty} {}_s \tilde{R}_{\ell m n} (r) = 1
\end{equation}
We can expand ${}_s \tilde{R}_{\ell m n}(r)$ as a series of Chebyshev polynomials and get ${}_s S_{\ell m n} (\theta)$ in terms of spin-weighted spherical functions as in Ref.~\cite{Cook:2014cta}. As a result, we can use our numerical results and evaluate them for any point $(t,r, \theta, \phi)$ of spacetime in the subsequent ray tracing section. To ensure the perturbation metric is flawless, we also verified $\hat{\mathcal{E}}_{ab}(h) = 0$ numerically.

\subsection{Radiated energy}

In this section, we estimate the radiated energy of a single mode after truncation. The energy flux is given by \cite{Berti:2007fi}
\begin{equation}
\dv{E}{t}=\lim _{r \rightarrow \infty}\left[\frac{r^2}{16 \pi} \int_{\Omega}\left|\int^t_{+\infty} \Psi_4^{(1)} d \tilde{t}\right|^2 \dd\Omega\right] \,, \\
\end{equation}
Then the total energy emitted can be calculated as
\begin{equation}
E_{\text{tot}} = \int_{t_0}^\infty \dv{E}{t} \dd{t}
\end{equation}
where $t_0 = r_*(r_o)$. This total energy satisfies the condition $E_{\mathrm{tot}} / M \lessapprox 3\%$ \cite{Berti:2007fi}. Assuming that only the $\ell = m = 2$ mode dominates the radiated energy, this relationship leads to
\begin{equation}
|{}_{s}Z| \lessapprox 0.24 \,.
\end{equation}
It is important to note that this approach differs slightly from truncating the metric using $H(t - r_*)$. However, it should provide a sufficiently accurate estimation for our purposes.

\section{Backward Ray Tracing}

\subsection{Local oberver basis}
\label{sec:observer}

The observation of photons at any given point is dependent on the observer. For the case of a pure Kerr black hole, zero angular momentum observers (ZAMOs) are typically employed.
In this section, we extend them to dynamically perturbed Kerr spacetime.

Within the Boyer-Lindquist coordinate system, a ZAMO frame can be chosen and expressed in terms of the coordinate basis $\left\{\partial_t, \partial_r, \partial_\theta, \partial_\phi\right\}$ \cite{Cunha:2016bpi, Frolov:1998wf, Johannsen:2013vgc, Bardeen:1973tla} as
\be
\vec{z}_{(t)}^\mu  = \frac{g_{\phi\phi} \partial_t - g_{\phi t} \partial_\phi}{\sqrt{g_{\phi \phi}\left(g_{\phi t}^2-g_{\phi \phi} g_{t t}\right)}} \,,\, \vec{z}_{(r)}^\mu = \frac{\partial_r}{\sqrt{g_{rr}}} \,,\,
\vec{z}_{(\theta)}^\mu =\frac{ \partial_{\theta}}{\sqrt{g_{\theta \theta}}} \,, \,\vec{z}_{(\phi)}^\mu = \frac{\partial_\phi}{\sqrt{g_{\phi\phi}}} \,.
\ee
The observer basis $\vec{z}_{(\sigma)}^\nu$ has a Minkowski normalization
\begin{equation}
g^{(0)}_{\mu\nu} \vec{z}_{(\rho)}^\mu \vec{z}_{(\sigma)}^\nu = \eta_{(\rho)(\sigma)} \,, \label{eq:mink_norm}
\end{equation}
with $\eta_{(\rho)(\sigma)} = \diag\{-1,1,1,1\}$.
Unlike the unperturbed case, Eq.~\eqref{eq:mink_norm} no longer holds when the unperturbed metric $g^{(0)}_{\mu\nu}$ is replaced by the perturbed metric $g_{\mu\nu}$. 
To address this issue, we employ a Gram-Schmidt orthogonalization process, using the chosen ZAMO tetrad $\vec{v}_\rho = \vec{z}_{(\rho)}^\mu$ as the initial set of vectors.
The Gram-Schmidt process can be expressed as
\begin{equation}
\begin{aligned}
\vec{u}_0 & =\vec{v}_0 \,, \\
\vec{u}_1 & =\vec{v}_1-\operatorname{proj}_{\vec{u}_0}\left(\vec{v}_1\right) \,, \\
\vec{u}_2 & =\vec{v}_2-\operatorname{proj}_{\vec{u}_0}\left(\vec{v}_2\right)-\operatorname{proj}_{\vec{u}_1}\left(\vec{v}_2\right) \,, \\
\vec{u}_3 & =\vec{v}_3-\operatorname{proj}_{\vec{u}_0}\left(\vec{v}_3\right)-\operatorname{proj}_{\vec{u}_1}\left(\vec{v}_3\right)-\operatorname{proj}_{\vec{u}_2}\left(\vec{v}_3\right) \,, 
\end{aligned}
\end{equation}
where the projection operator is defined as
\begin{equation}
\operatorname{proj}_{\vec{u}} \vec{v}=\frac{\langle\vec{u}, \vec{v}\rangle}{\langle\vec{u}, \vec{u}\rangle} \vec{u} \,,
\end{equation}
with the inner product evaluated using the metric $g_{\mu\nu}$.
The resulting orthonormalized tetrad, denoted as
\begin{equation}
\tilde{\vec{z}}_{(\rho)}^\mu = \frac{\vec{u}_\rho}{\sqrt{\abs{\langle\vec{u}_\rho \,, \vec{u}_\rho\rangle}}}
\end{equation}
satisfies the following condition
\begin{equation}
g_{\mu\nu} \tilde{\vec{z}}_{(\rho)}^\mu \tilde{\vec{z}}_{(\sigma)}^\nu = \eta_{(\rho)(\sigma)} \,. \label{eq:mink_norm_perturbed}
\end{equation}

\subsection{Impact Parameters}

In the image plane of the observer defined above, each photon is assigned Cartesian coordinates $(\tilde{X}, \tilde{Y})$, which represent its impact parameters. These coordinates are proportional to the photon's observation angles $(\tilde\alpha, \tilde\beta)$. Here we follow a convention similar to Ref.~\cite{Cunha:2016bpi}, and define them as:

\begin{equation}
\tilde{X} \equiv -r \tilde\beta, \quad \tilde{Y} \equiv r \tilde\alpha \,.
\end{equation}
Considering the geometry of photon detection, we can express the photon's 4-momentum in the observer's reference frame as follows (refer to Fig.~1 in Ref.~\cite{Cunha:2016bpi}):
\be
p^{(\phi)} = |\vec{P}| \sin \tilde\beta \cos \tilde\alpha \,,\,
p^{(\theta)} = |\vec{P}| \sin \tilde\alpha \,,\,
p^{(r)} = |\vec{P}| \cos \tilde\beta \cos \tilde\alpha \,,
\ee
where $|\vec{P}| = p^{(t)}$. From this, we can derive the photon's momentum using
\begin{equation}
p^\mu = \tilde{\vec{z}}_{(\sigma)}^\mu p^{(\sigma)} \,.
\end{equation}

\subsection{Kerr–Schild coordinates}\label{sec:kerr_schild}
To avoid the numerical problem when a photon gets too close to the pole in the BL coordinate system, we perform ray tracing in the Cartesian Kerr–Schild (KS) coordinate system $\{t_{\mathrm{KS}}, x, y, z\}$. The coordinate transformations, which can be found in Ref.~\cite{Christian:2020xrp, Madler:2018tkl, 10.1046/j.1365-8711.1999.02459.x}, are also reviewed in the following.

The Kerr metric in Kerr–Schild coordinates $\{t_{\mathrm{KS}}, x, y, z\}$ can be expressed as
\begin{equation}
\begin{aligned}
& g_{\mu \nu}^{(0)}=\eta_{\mu \nu}+f k_\mu k_\nu \,, \\
& f=\frac{2 M r^3}{r^4+a^2 z^2} \,,\, k_\mu = \left(1, \frac{r x+a y}{r^2+a^2}, \frac{r y-a x}{r^2+a^2}, \frac{z}{r}\right) \,,
\end{aligned}
\end{equation}
where $r$ is the same as the radial coordinate in the BL coordinates. The radial coordinate $r$ can be defined in terms of the Cartesian KS coordinates as
\begin{equation}
\frac{x^2+y^2}{r^2+a^2}+\frac{z^2}{r^2}=1 \,.
\end{equation}
Then the coordinates $\{r, \theta, \phi_{\mathrm{KS}}\}$ can be written explicitly in terms of Cartesian KS as
\begin{equation}
\begin{aligned}
r &= \left[\frac{R^2-a^2+\sqrt{\left(R^2-a^2\right)^2+4 a^2 z^2}}{2}\right]^{1 / 2}, \\
\theta &= \arccos{\frac{z}{r}} \,, \,
\phi_{\mathrm{KS}} = \tan^{-1}(-a y-r x,a x-r y)+\pi \,.
\end{aligned}
\end{equation}
where
\begin{equation}
R = \sqrt{x^2 + y^2 + z^2} \,.
\end{equation}
The function $\tan^{-1}(x, y)$, also known as the two-argument arctangent function, calculates the arctangent of $y/x$ while considering the quadrant in which the point $(x, y)$ lies.

The coordinate transformation from BL coordinates can be written as
\begin{equation}
\begin{aligned}
t_{\mathrm{KS}} & = t + \int \frac{2 M r}{\Delta_{\mathrm{BL}}} dr \,,\\
x & =\sqrt{r^2+a^2} \sin \theta \cos\left[\phi_{\mathrm{KS}} + \arctan (a / r)\right] \,,\\
y & =\sqrt{r^2+a^2} \sin \theta \sin \left[\phi_{\mathrm{KS}} + \arctan (a / r)\right] \,,\\
z & =r \cos \theta \,,
\end{aligned}
\end{equation}
where
\begin{equation}
\phi_{\mathrm{KS}} = \phi + a \int \frac{d r}{\Delta_{\mathrm{BL}}}\,.
\end{equation}
Following Ref.~\cite{Christian:2020xrp}, we fix the integration constant is fixed by letting $t_{\mathrm{KS}} = t$ and $\phi_{\mathrm{KS}} = \phi$ at the initial radius $r_o$, so we have
\begin{equation}
\begin{aligned}
t_{\mathrm{KS}} & = t + \frac{M}{\sqrt{M^2 - a^2}}\left[ r_{+} \ln \left(\frac{r-r_{+}}{r_o-r_{+}}\right)-r_{-}\ln \left(\frac{r-r_{-}}{r_o-r_{-}}\right)\right] \,, \\
\phi_{\mathrm{KS}} & = \phi + \frac{a}{2 \sqrt{M^2-a^2}}\left[\ln \left(\frac{r-r_{+}}{r_o-r_{+}}\right)-\ln \left(\frac{r-r_{-}}{r_o-r_{-}}\right)\right] \,.
\end{aligned}
\end{equation}

\section{Results of Mirror Mode}

\begin{figure}[h]
   \begin{minipage}{0.48\textwidth}
     \centering
     \includegraphics[width=\linewidth]{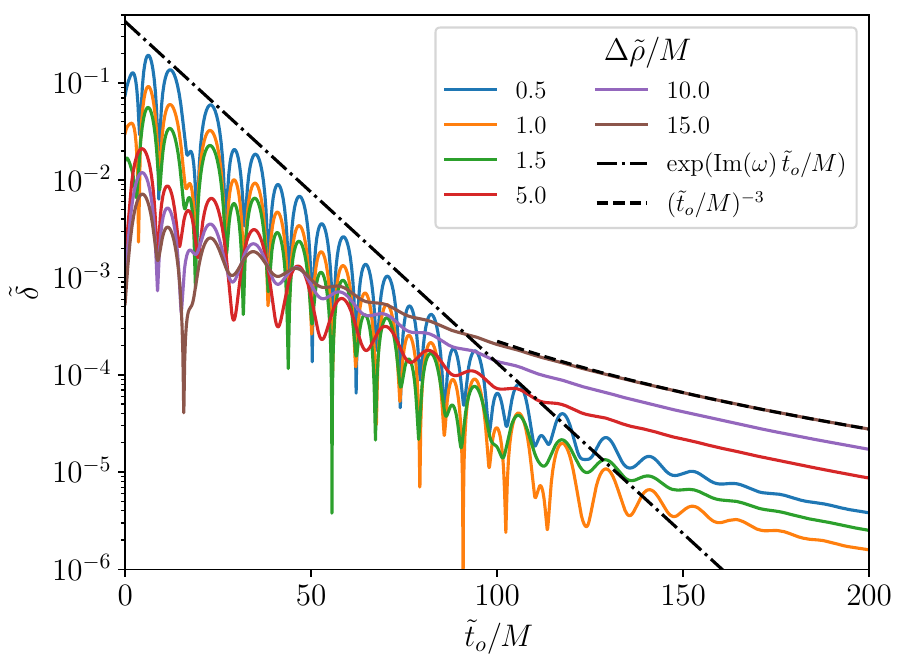}
   \end{minipage}
   \hfill
   \begin{minipage}{0.48\textwidth}
     \centering
     \includegraphics[width=\linewidth]{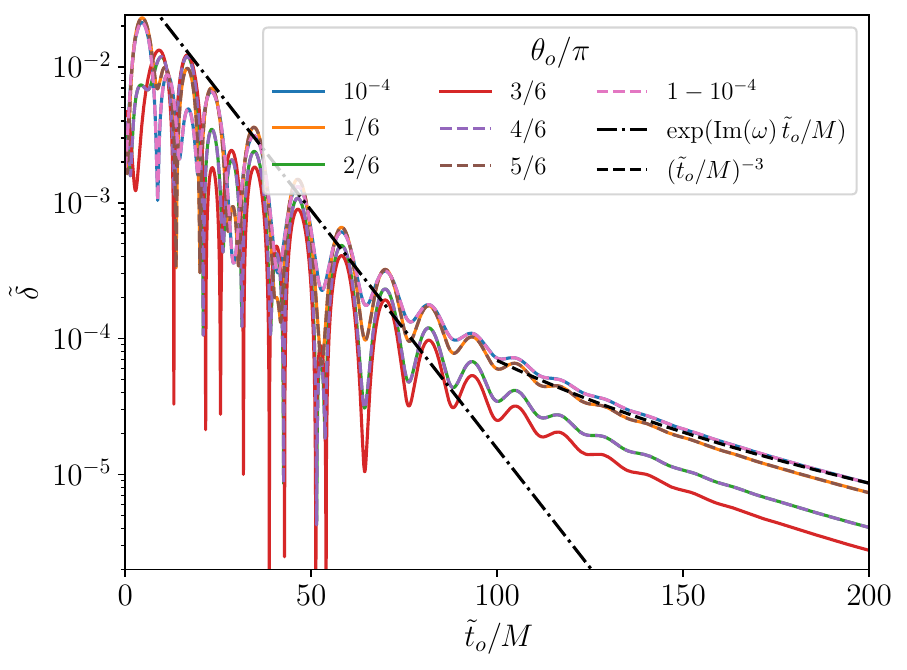}
   \end{minipage}
   \caption{Evolution of the deflection angles of geodesics terminating at infinity, plotted as a function of observation time $\tilde{t}_o$, for the case incorporating the QNM's mirror mode with equal energy.
   \textbf{Left:} Results for various impact parameters $\Delta \tilde{\rho}/M$ with $\theta_o / \pi = 1 - 10^{-4}$. \textbf{Right:} Results for various initial inclination angles $\theta_0$ with $\tilde{\rho} / M = 10$ along the $+\tilde{X}$ axis.}   \label{fig:mirror}
\end{figure}

In this section, we consider the effect of incorporating the QNM's mirror mode with equal energy, reflecting a more realistic scenario where the SMBHB was in a quasi-circular orbit before the merger. In Fig.~\ref{fig:mirror}, we show the evolution of $\tilde{\delta}$ for different values of $\Delta\tilde{\rho}$ and $\theta_o$. Compared to the results of Figs.~4 and 5 in the maintext, this setup also exhibits the two previously discussed characteristics: exhibit the same two key features: an initial rapid decay during the ringdown phase, followed by a late-time power-law tail. Due to the mirror mode's symmetry across the equatorial plane, the dependence on $\theta_o$ becomes less pronounced.

\end{document}